\documentclass[runningheads]{llncs}

\usepackage[utf8]{inputenc}
\usepackage{amsmath}
\usepackage{amsfonts}
\usepackage{graphicx}
\usepackage{color}
\usepackage[nocompress]{cite}

\usepackage{hyperref}

\usepackage{algorithm}
\usepackage{algorithmic}

\newcommand{\mat}[1]{\boldsymbol{\mathrm{#1}}}
\newcommand{\vecb}[1]{\boldsymbol{#1}}

\newcommand{\diff}{\mathop{}\!\mathrm{d}}

\DeclareMathOperator{\diag}{diag}
\DeclareMathOperator{\supp}{supp}

\begin{document}

\title{A fast Metropolis-Hastings method for generating random correlation
matrices}

\titlerunning{Uniform sampling of correlation matrices}

\author{Irene Córdoba\inst{1}\orcidID{0000-0002-3252-4234} \and Gherardo
Varando\inst{1,2}\orcidID{0000-0002-6708-1103} \and Concha
Bielza\inst{1}\orcidID{0000-0001-7109-2668} \and Pedro
Larrañaga\inst{1}\orcidID{0000-0003-0652-9872}}
\institute{Department of Artificial Intelligence, Universidad Politécnica de
Madrid \and Department of Mathematical Sciences, University of Copenhagen}

\authorrunning{I. Córdoba et al.}

\maketitle

\begin{abstract}
	We propose a novel Metropolis-Hastings algorithm to sample uniformly from
	the space of correlation matrices. Existing methods in the literature are
	based on elaborated representations of a correlation matrix, or on complex
	parametrizations of it. By contrast, our method is intuitive and simple,
	based the classical Cholesky factorization of a positive definite matrix and
	Markov chain Monte Carlo theory. We perform a detailed convergence analysis
	of the resulting Markov chain, and show how it benefits from fast
	convergence, both theoretically and empirically. Furthermore, in numerical
	experiments our algorithm is shown to be significantly faster than the
	current alternative approaches, thanks to its simple yet principled
	approach.
	\keywords{Correlation matrices \and Random sampling \and Metroplis-Hastings}
\end{abstract}

\section{Introduction}\label{sec:int}
Correlation matrices are a fundamental tool in statistics and for the analysis
of multivariate data. In many application domains, such as signal processing or
regression analysis, there is a natural need for tools that generate synthetic,
benchmark correlation matrices \cite{marsaglia1984,holmes1991,fallat2017}.
Existing algorithms for this task usually randomly sample either the eigenvalues
of the matrix or the elements of its Cholesky decomposition, instead of directly
sampling the matrix from its uniform distribution.

Uniform sampling of correlation matrices has received little attention until
recently \cite{pourahmadi2015,lewandowski2009}. By contrast with the classical methods
\cite{marsaglia1984,holmes1991}, uniform sampling does not assume any a priori
information and allows to obtain an unbiased random correlation matrix. In this
paper, we propose a new Metropolis-Hastings algorithm for such task, which is
significantly faster than the existing algorithms in the literature
\cite{lewandowski2009,pourahmadi2015}. We perform a detailed analysis of the
convergence properties of the Markov chain that we construct.

Our approach is similar to that of Pourahmadi and Wang~\cite{pourahmadi2015} in
the sense that we rely on the Cholesky factorization of a positive definite
matrix; however they reparametrize
the triangular factor with spherical coordinates, resulting in an additional layer
of complexity. Lewandowsky et al.~\cite{lewandowski2009} present two methods
based on alternative representations of the correlation matrix, vines and
elliptical distributions, which are arguably less direct than our classical
Cholesky factorization.

The rest of the paper is organized as follows. In Section \ref{sec:prel} we
briefly overview the Cholesky factorization of correlation matrices, and other
technical results needed for our method. Section \ref{sec:mcmc} contains the
details of our Metropolis-Hastings algorithm, whose convergence properties are
analyzed in Section \ref{sec:conv}, both from a theoretical and experimental
point of view. In Section \ref{sec:perf} we empirically compare the
computational performance of our method with the alternatives in the literature.
Finally, we conclude the paper in Section \ref{sec:conc}.

\section{Upper Cholesky factorization of a correlation matrix}\label{sec:prel}
Let $\mat{R}$ be $p \times p$ a correlation matrix, that is, a symmetric positive
definite~(SPD) matrix with ones on the diagonal.  Since $\mat{R}$ is SPD, 
it has a unique upper Cholesky factorization
$\mat{R}=\mat{U}\mat{U}^t$, with $\mat{U}$  an upper  triangular matrix
with positive diagonal entries.   
Let $\mathcal{U}$ denote the set of upper triangular $p \times p$ matrices with positive
diagonal entries. We will define the set of SPD correlation matrices as
\begin{equation} \label{eq:corrparam}
    \mathcal{R} = \{\mat{R} = \mat{U}\mat{U}^t \text{ s.t. } \diag(\mat{R}) =
	\vecb{1} \text{, } \mat{U} \in \mathcal{U}\}.
\end{equation}

The set $\mathcal{R}$ of SPD correlation matrices is known to form a convex body
called \emph{elliptope} \cite{laurent1996}, whose volume has been explicitly
computed by Lewandowski et al.~\cite{lewandowski2009}. 
Observe that the constraint $\diag(\mat{R}) =
\vecb{1}$ in Equation~\eqref{eq:corrparam} simply translates
to the rows of $\mat{U}$ being normalized vectors. 
Denoting the subset of $\mathcal{U}$ with such normalized rows as
$\mathcal{U}_1$, $\mathcal{R}$ can be written more compactly as
\[
	\mathcal{R} = \{ \mat{R} = \mat{U}\mat{U}^t \text{ s.t. } \mat{U} \in
	\mathcal{U}_1\}.
\]

Consider now $\Phi(\mat{U}) = \mat{U} \mat{U}^{t}$ as a parametrization of
$\mathcal{U}_1$ into SPD matrices. In order to
sample uniformly from $\mathcal{R}$, which is the image of $\Phi$, we need to
compute the Jacobian matrix
$J \Phi(\mat{U})$ \cite{diaconis2013}. Then, when sampling in $\mathcal{U}_1$ from a density
proportional to the Jacobian $\det(J \Phi(\mat{U}))$, the induced distribution on
$\mathcal{R}$ by $\Phi$ is the uniform measure.
In our case, the Jacobian is~\cite{eaton1983}
\begin{equation}\label{eq:jacob}
	\det(J\Phi(\mat{U})) = 2^{p}\prod_{i = 1}^{p - 1} u_{ii}^i,
\end{equation}
where $u_{ii}$ is the $i$-th diagonal element of $\mat{U} \in \mathcal{U}_1$ and
we have omitted $u_{pp}$ because it is equal to $1$.

\section{Metropolis-Hastings uniform sampling}\label{sec:mcmc}
We will use a Metropolis-Hastings method for sampling from a density
proportional to the Jacobian
in Equation \eqref{eq:jacob}. Observe that the $i$-th row in $\mat{U}$,
denoted as $\vecb{u}_i$ in the remainder, can be
sampled independently from all the other rows $\{\vecb{u}_j\}_{i\neq j}$, from a 
density $f(\vecb{u}_i) \propto u_{ii}^i$.
Furthermore, $\vecb{u}_i$ is a unitary vector and has its first $i - 1$ entries
equal to zero, therefore it lives in the $(p - i)$-dimensional hemisphere,
\[
	\mathcal{S}^{p - i}_+ = \{\vecb{v} \in \mathbb{R}^{p - i + 1} \text{ s.t. }
	\vecb{v}\vecb{v}^t = 1 \text{ and } v_1 > 0 \},
\]
where the positivity constraint is to ensure that $u_{ii} > 0$.
This independent row-wise sampling procedure is described in Algorithm
\ref{alg:mhfull}.
\begin{algorithm}
	\caption{Uniform sampling in $\mathcal{R}$}
\label{alg:mhfull}
\begin{algorithmic}[1]
	\setlength{\lineskip}{0.13cm}

	\REQUIRE Sample size $N$ 
	\ENSURE Uniform sample from $\mathcal{R}$ of size $N$ 
	
	\FOR {$n = 1, \ldots, N$}
		\STATE $\mat{U}^n \gets \mat{0}_p$
		\FOR {$i = 1, \ldots, p$}
		\STATE $\vecb{u}^n_{i} \gets$ sample from $f(\vecb{u}_i) \propto
		u_{ii}^i$ on $\mathcal{S}_+^{p-i}$\label{alg:l:rowwise}
		\ENDFOR
	\ENDFOR

	\RETURN $\{\Phi(\mat{U}^1), \ldots, \Phi(\mat{U}^N)\}$
	\end{algorithmic}
\end{algorithm}

Since each row of $\mat{U}$ can be sampled independently, in the remainder of this
section we will concentrate on how to perform step \ref{alg:l:rowwise} in
Algorithm \ref{alg:mhfull}. In order to lighten the notation, we will restate
our problem as sampling vectors $\vecb{v}$
from the hemisphere $\mathcal{S}^{p - i}_+$ with respect to the density
$f(\vecb{v}) \propto v_1^i$, where $p$ is fixed and $1 \leq i < p$.  

In the Metropolis-Hastings algorithm, we need to generate a proposed vector
$\tilde{\vecb{v}}$ from the current vector $\vecb{v}$ already sampled from
$\mathcal{S}^{p - i}_+$. For this, we will propose the new state as a
normalized perturbation of the current vector, specifically,
\begin{equation}\label{eq:prop}
	\tilde{\vecb{v}} = \frac{\vecb{v} +
	\vecb{\epsilon}}{\lVert \vecb{v} + \vecb{\epsilon}
	\rVert},
\end{equation}
where $\vecb{\epsilon}$ is a Gaussian random vector of dimension $p - i + 1$
with zero mean and component-independent variance $\sigma_\epsilon^2$. 

With the transformation of Equation \eqref{eq:prop} the induced proposal distribution $q(\tilde{\vecb{v}}
| \vecb{v})$ is a projected Gaussian over $\mathcal{S}^{p - i}_+$ \cite{mardia1999},
with parameters $\vecb{v}$ and $\sigma^2_\epsilon \mat{I}_{p - i + 1}$.
The expression for the density of this angular distribution is given in the
general case by \cite{pukkila1988}. In our setting, we obtain a simplified
expression, 
\begin{equation}\label{eq:prop:dens}
	q(\tilde{\vecb{v}} | \vecb{v}) = \frac{\exp\left((
			(\vecb{v}^t\tilde{\vecb{v}})^2 -1
	)/2\sigma^2_\epsilon\right)}{(2\pi)^{(p - i + 1)/2}}
	\int_0^\infty
s^{p-i}\exp\left(-\frac{1}{2}\left(s - \frac{\vecb{v}^t\tilde{\vecb{v}}
	}{\sigma_\epsilon}\right)^2\right)
\diff s.
\end{equation}
The density for the proposal
$q(\tilde{\vecb{v}} | \vecb{v})$ in Equation \eqref{eq:prop:dens} is a function
of the scalar product $\vecb{v}^t\tilde{\vecb{v}}$, therefore it is symmetric because the roles of
$\vecb{v}$ and $\tilde{\vecb{v}}$ can be exchanged, and we can omit the Hastings
correction from the sampling scheme.
Thus the acceptance probability at each step of the algorithm becomes
\[
	\min\left(1, \frac{f(\tilde{\vecb{v}})}{f(\vecb{v})}\right) = 
        \min\left(1, \mathbb{I}_{\geq
        0}(\tilde{v}_1)\left(\frac{\tilde{v}_{1}}{v_1}\right)^i\right),
\]
where $\tilde{v}_1$ is the first component of the proposed vector
$\tilde{\vecb{v}}$ and $\mathbb{I}_{\geq 0}$ denotes the indicator function of
the positive real numbers.
The described Metropolis sampling is illustrated in Algorithm \ref{alg:mhrow}.
\begin{algorithm}
	\caption{Metropolis sampling of vectors $\vecb{v}$ in $\mathcal{S}_+^{p-i}$
	from $f(\vecb{v}) \propto v_1^i$}
\label{alg:mhrow}
\begin{algorithmic}[1]
	\setlength{\lineskip}{0.13cm}

	\REQUIRE Sample size $N$, variance $\sigma_\epsilon^2$
	 and burn-in time $t_{b}$ 
	\ENSURE Sample of size $N$ from $S_+^{p-i}$ 
	
	\STATE $\vecb{v}_0 \gets$ random standard multivariate Gaussian observation of dimension $p -
	i + 1$
	\STATE $v_{01} \gets | v_{01} |$
	\STATE $\vecb{v}_0 \gets$ normalize $\vecb{v}_0$

	\FOR {$t = 0, \ldots, t_b + N$}
		\FOR {$j = 1, \ldots, p - i + 1$}
			\STATE $\epsilon_j \gets$ random Gaussian observation with zero mean
                        and variance 
			$\sigma_\epsilon^2$
		\ENDFOR
		\STATE $\tilde{\vecb{v}} \gets \vecb{v}_t +
		\vecb{\epsilon}$
		\STATE $\tilde{\vecb{v}} \gets$ normalize $\tilde{\vecb{v}}$
		\STATE $\delta \gets$ random uniform observation on $[0, 1]$
		\IF {$\tilde{v}_1 \geq 0$ \AND $\delta \leq (\tilde{v}_1/v_{t1})^i$}
			\STATE $\vecb{v}_t \gets \tilde{\vecb{v}}$
		\ENDIF
	\ENDFOR

	\RETURN $\{\vecb{v}_{t_{b} + 1}, \vecb{v}_{t_{b} + 2},
	, \ldots, \vecb{v}_{t_{b} + N}\}$
	\end{algorithmic}
\end{algorithm}

\section{Convergence assessment}\label{sec:conv}
In this section we will analyze, both theoretically and empirically, the
convergence properties of the proposed Algorithm \ref{alg:mhrow}, following
\cite{robert2004}. 

\subsection{Theoretical convergence properties}
A minimal requirement for a Metropolis chain with proposal $q$ to have the target density $f$ as
its stationary distribution is the following relationship between the supports
\[
	\supp(f) \subseteq \bigcup_{\vecb{v} \in \supp(f)}\supp(q(\cdot \mid
	\vecb{v})).
\]
Since in our case the support of $q(\cdot \mid \vecb{v})$ is the $(p -
i)$-dimensional unit sphere, for all $\vecb{v} \in \mathcal{S}_+^{p-i}$, this
condition is automatically satisfied.  

Given the above minimal requirement, if the
chain additionally is $f$-irreducible and aperiodic, then it converges to its
stationary distribution (Theorem 7.4 in \cite{robert2004}). The fist condition
holds in our case because the proposal is strictly positive for all $\vecb{v},
\tilde{\vecb{v}} \in \mathcal{S}_+^{p-1}$. A sufficient
condition for aperiodicity is that the probability of remaining in the same
state for the next step is strictly positive, that is, $P(f(\vecb{v}) \geq
f(\tilde{\vecb{v}})) > 0$. In our case, we have 
\begin{equation*}
		P(f(\vecb{v}) \geq f(\tilde{\vecb{v}})) = P\left(v_1^i \geq 
\mathbb{I}_{\geq 0}(\tilde{v}_1)\tilde{v}_1^i\right) 
		= 
			P\left(v_1 \geq \tilde{v}_1, \tilde{v}_1 \geq 0\right)  +
			P\left(\tilde{v}_1  \leq 0\right).
\end{equation*}
Expanding the second summand and using the fact that $v_1 \geq 0$, we obtain
\begin{equation}\label{eq:preject}
		P (\tilde{v}_1 \leq 0) = P(\epsilon_1 \leq 0) - P(-v_1 \leq \epsilon_1 \leq 0)
		=
			\frac{1}{2} - \int_0^{v_1}
			\frac{1}{\sqrt{2\pi\sigma_\epsilon}}e^{-{s^2}/(2\sigma_\epsilon^2)}\diff
			s,
\end{equation}
Therefore $P(f(\vecb{v}) \geq f(\tilde{\vecb{v}}))$ is strictly positive, the
chain is aperiodic, and our proposed algorithm converges to $f$. 

Some additional insights can be gained on the convergence of the algorithm when
the variance $\sigma_\epsilon^2$ increases. From Equation
\eqref{eq:prop:dens}, we observe that in such scenario, the proposal density
approaches to a
constant, yielding an independent Metropolis algorithm \cite{robert2004}. 
The expression for such limiting proposal, which coincides with the inverse
sphere volume, is
\begin{equation}\label{eq:unifprop}
	\lim_{\sigma_\epsilon \to \infty} q(\tilde{\vecb{v}} \mid \vecb{v}) =
	\frac{\Gamma((p - i + 1)/2)}{2\pi^{(p - i + 1)/2}}.
\end{equation}
In this scenario, denoting as $C_f$ and $C_q$ the integration constants for
$f$ and $q$ respectively, taking $M \geq
(C_qC_f)^{-1}$ we have for all $\vecb{v} \in
\mathcal{S}_{+}^{p-i}$ that $f(\vecb{v}) \leq Mq(\vecb{v})$. Therefore the chain
is uniformly ergodic, and $2(1 - M^{-1})^n$ is an upper bound for the total variation norm
between the transition kernel after $n$ iterations and the target
distribution $f$ (Theorem 7.8 in \cite{robert2004}). Furthermore, $M^{-1}$ is
also a lower bound for the expected acceptance probability. 

\subsection{Empirical monitoring}
The above theoretical analysis assures the convergence of the proposed Algorithm
\ref{alg:mhrow}. Such convergence can also be empirically monitored in order to
get insight on how to tune its hyper-parameters: the burn-in time $t_b$ and the
perturbation variance $\sigma_\epsilon^2$. This will be the focus of this
subsection. For this task, the most challenging matrices are arguably high
dimensional therefore we will focus on the case where $p = 1000$.  

There is no standard assessment scheme to follow that will guarantee an expected
behaviour for the Metropolis chain \cite{robert2004}. However, we can study the
behaviour of some characteristic quantities. We have chosen to focus on the
acceptance ratio, that is, the percentage of times that we have accepted the
proposed value, so it can be thought of as an approximation for $P(f(\vecb{v})
\leq f(\tilde{\vecb{v}}))$. 

\begin{figure}[h]
\centering
\includegraphics[width = 0.45\textwidth]{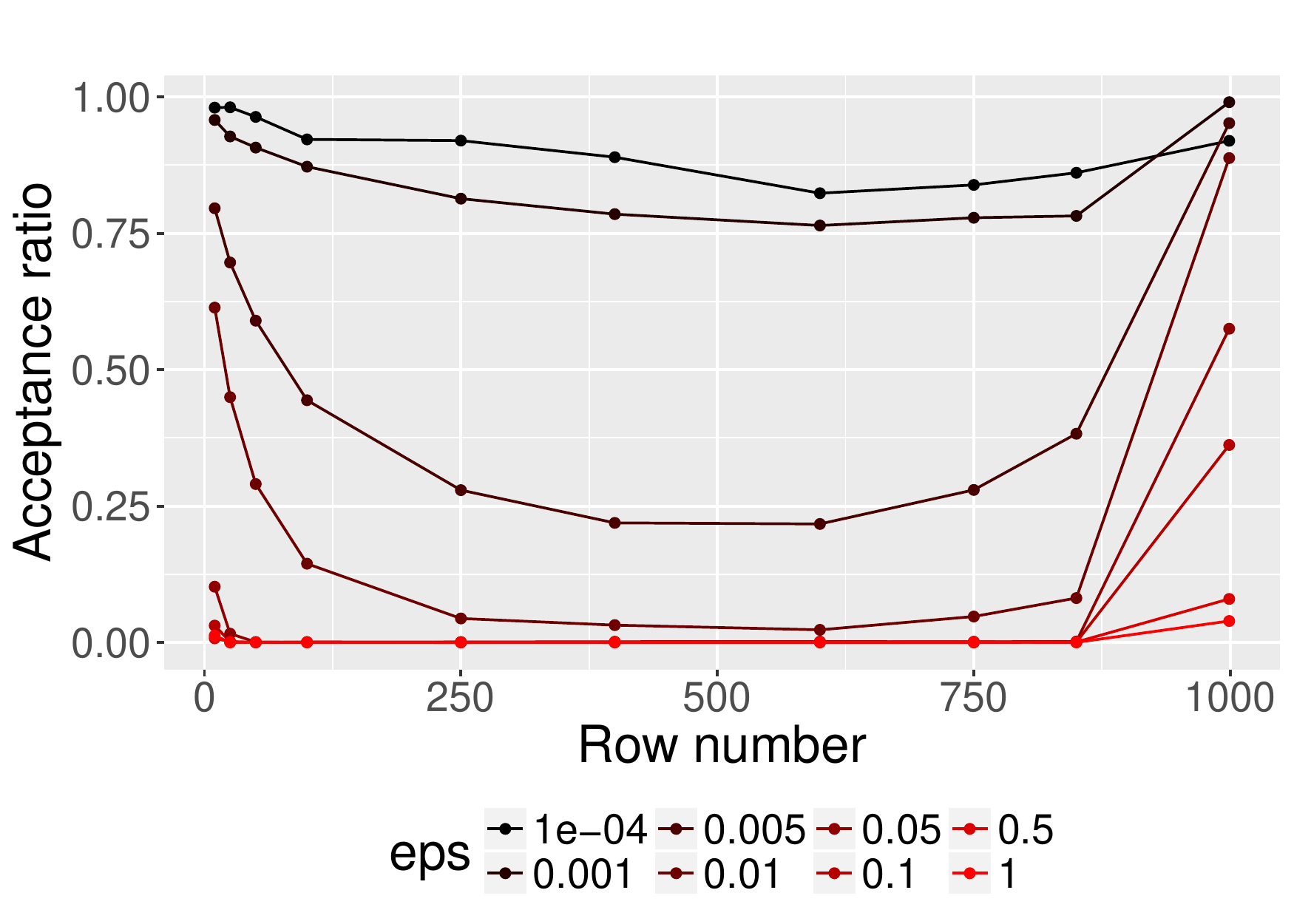} 
\includegraphics[width = 0.45\textwidth]{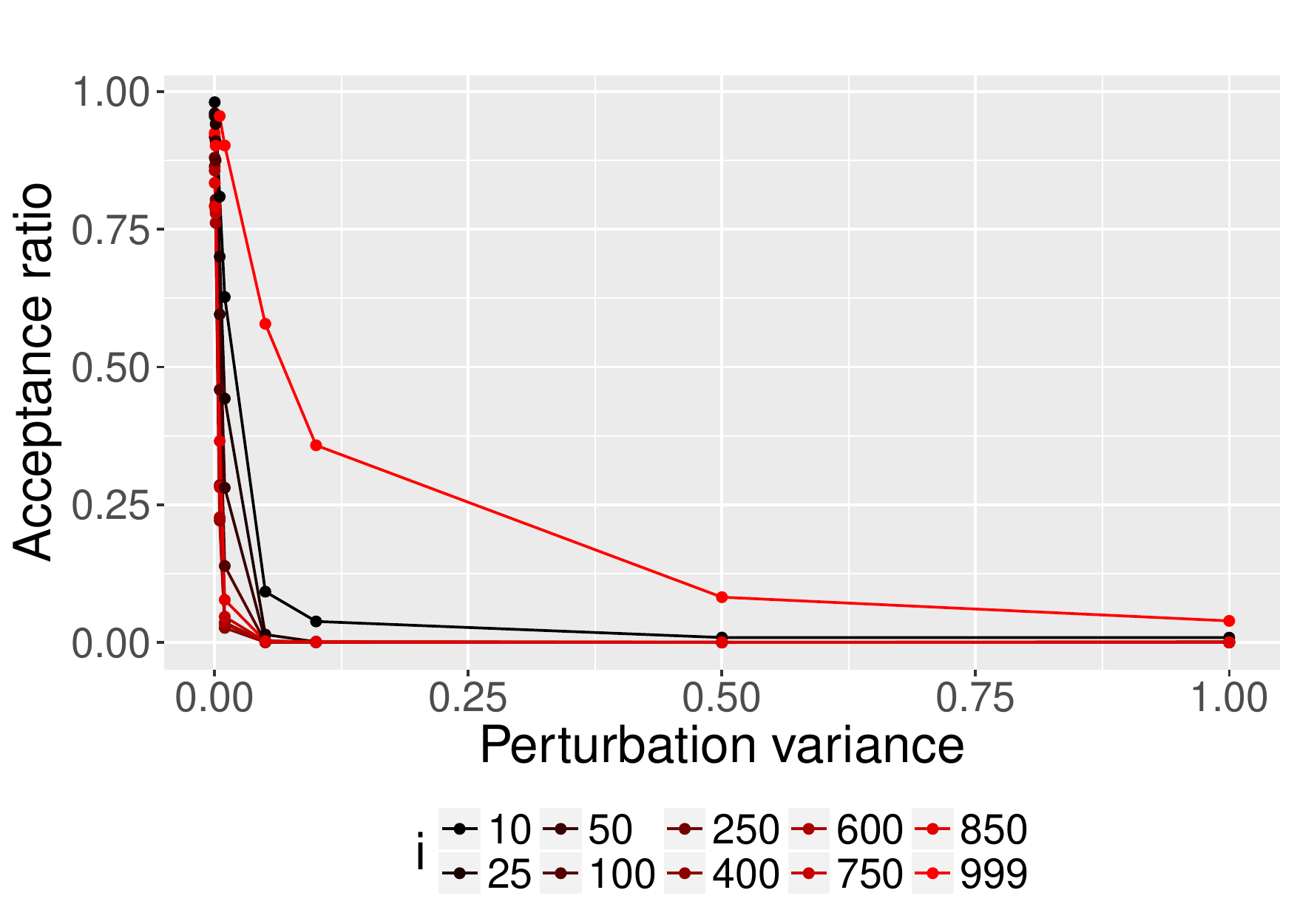} 
\caption{Acceptance ratio as a function of the row number $i$ (left) and the
perturbation variance $\sigma_\epsilon^2$ (right). eps: $\sigma_\epsilon^2$.}
\label{fig:acc}
\end{figure}

Whether a high acceptance ratio is desirable or not depends on the particular chain designed. 
For our case, in Figure \ref{fig:acc} this quantity is depicted as
a function of the row number and, complementarily, of the perturbation variance
$\sigma_\epsilon^2$.

We observe how, as $\sigma_\epsilon^2$ increases, the proposed value is rejected
more often. This could be already expected by looking at Equation
\eqref{eq:preject} above, where we see that the second term goes to zero as
$\sigma_\epsilon^2$ increases, yielding $\lim_{\sigma_\epsilon^2 \to \infty}
P(f(\vecb{v}) \leq f(\tilde{\vecb{v}})) \leq 1/2$. Furthermore, as
$\sigma_\epsilon$ increases the proposal distribution is more similar to the
uniform density on the $(p - i)$-dimensional sphere (Equation
\eqref{eq:unifprop}), which also hints the higher rejection rate. 

The row number $i$ also has a significant influence on the acceptance ratio.
Recall that $1 \leq i \leq p - 1$, $\vecb{v} \in \mathcal{S}^{p-i}_+$ and
$f(\vecb{v}) \propto v_1^i$, therefore as the row number $i$ increases the
target distribution $f$ approaches a delta function, and the dimensionality of
$\vecb{v}$ decreases.  Therefore it is reasonable to assume that the larger $i$
is, the smaller $\sigma_\epsilon^2$ should be for achieving a high acceptance
ratio, since it means that we are proposing new states that are, with high
probability, very close to the current state. 

\begin{figure}[h]
\centering
\includegraphics[width = 0.6\textwidth]{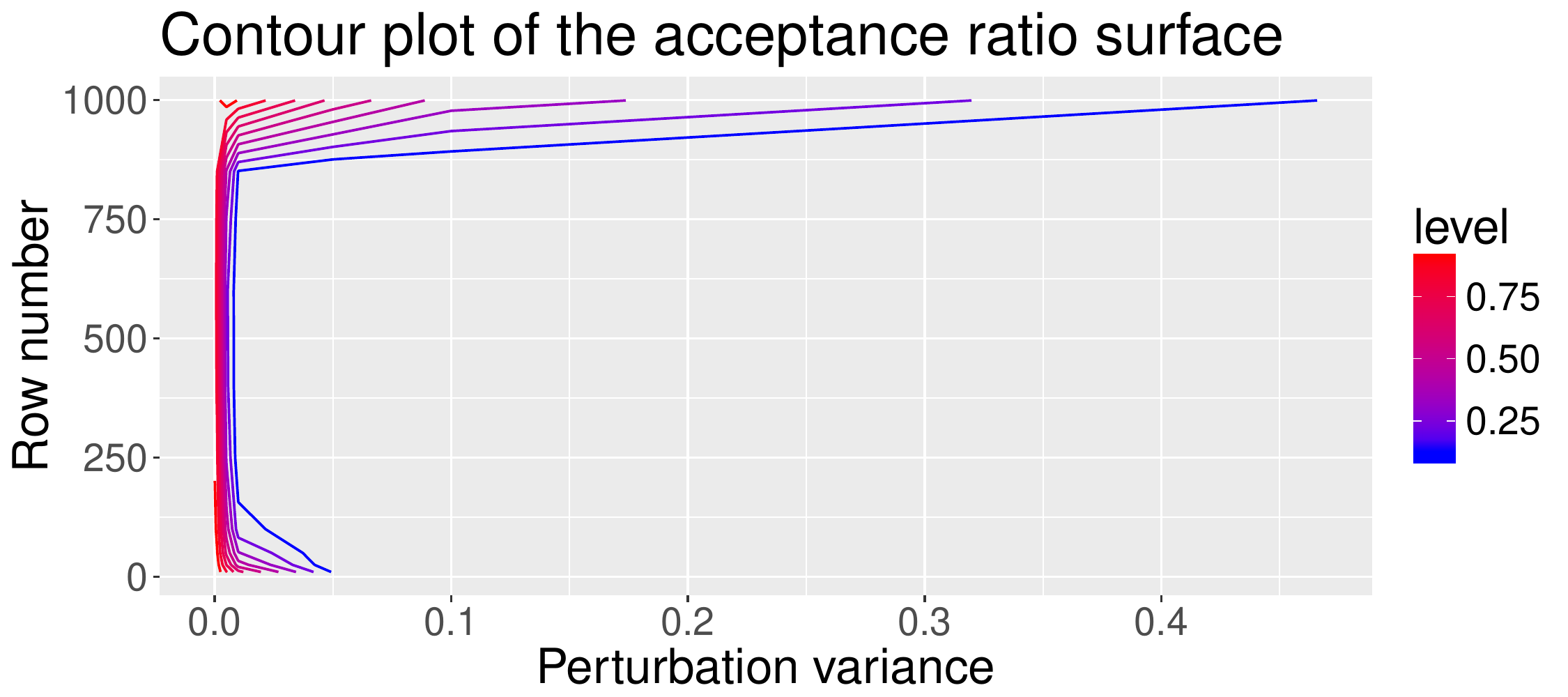} 
\caption{Contour lines of the acceptance ratio surface. level: magnitude of the
acceptance ratio.}
\label{fig:acc:3d}
\end{figure}

The above conclusions are further illustrated in Figure \ref{fig:acc:3d},
where we have plotted the contour lines of the acceptance ratio surface as a function of
$\sigma_\epsilon^2$ and the row number $i$. 
Observe that small values for $\sigma_\epsilon$ always lead to
high acceptance ratios, however this might not always be desirable since it  
can be a sign of slow convergence. By contrast, a low acceptance ratio can be
expected when approaching to a delta in moderately high dimensions, as is the case for row
numbers approximately between $250$ and $750$.

\section{Performance analysis}\label{sec:perf}
In this section we will compare our method, in terms of computational
performance, with the existing approaches in the literature for the same task:
uniform sampling of correlation matrices. We will generate $5000$ correlation
matrices of dimension $p = 10, 20, \ldots, 100$ using our algorithm, the vine
and onion methods of Lewandowski et al.~\cite{lewandowski2009}, and the polar
parametrization of Pourahmadi \cite{pourahmadi2015}.

Our algorithm has been implemented in R~\cite{rcore}. The vine and onion methods
are available in the function \emph{genPositiveDefMat} from the R package
\emph{clusterGeneration}\footnote{\url{https://CRAN.R-project.org/package=clusterGeneration}},
provided by the authors.  Since we have not found an implementation of the polar
parametrization method of Pourahmadi~\cite{pourahmadi2015}, we have developed
our own function, also in R, mimicking the method therein described. For our
method, based on the analysis of the previous section we have fixed
$\sigma_\epsilon = 0.01$ and $t_b = 1000$, which have empirically provided good
convergence results.  The experiment has been executed on a machine equipped
with Intel Core i7-5820k, $3.30$ GHz$\times12$ and $16$ GB of RAM.  

\begin{figure}[h]
	\centering
	\includegraphics[width = 0.45\textwidth]{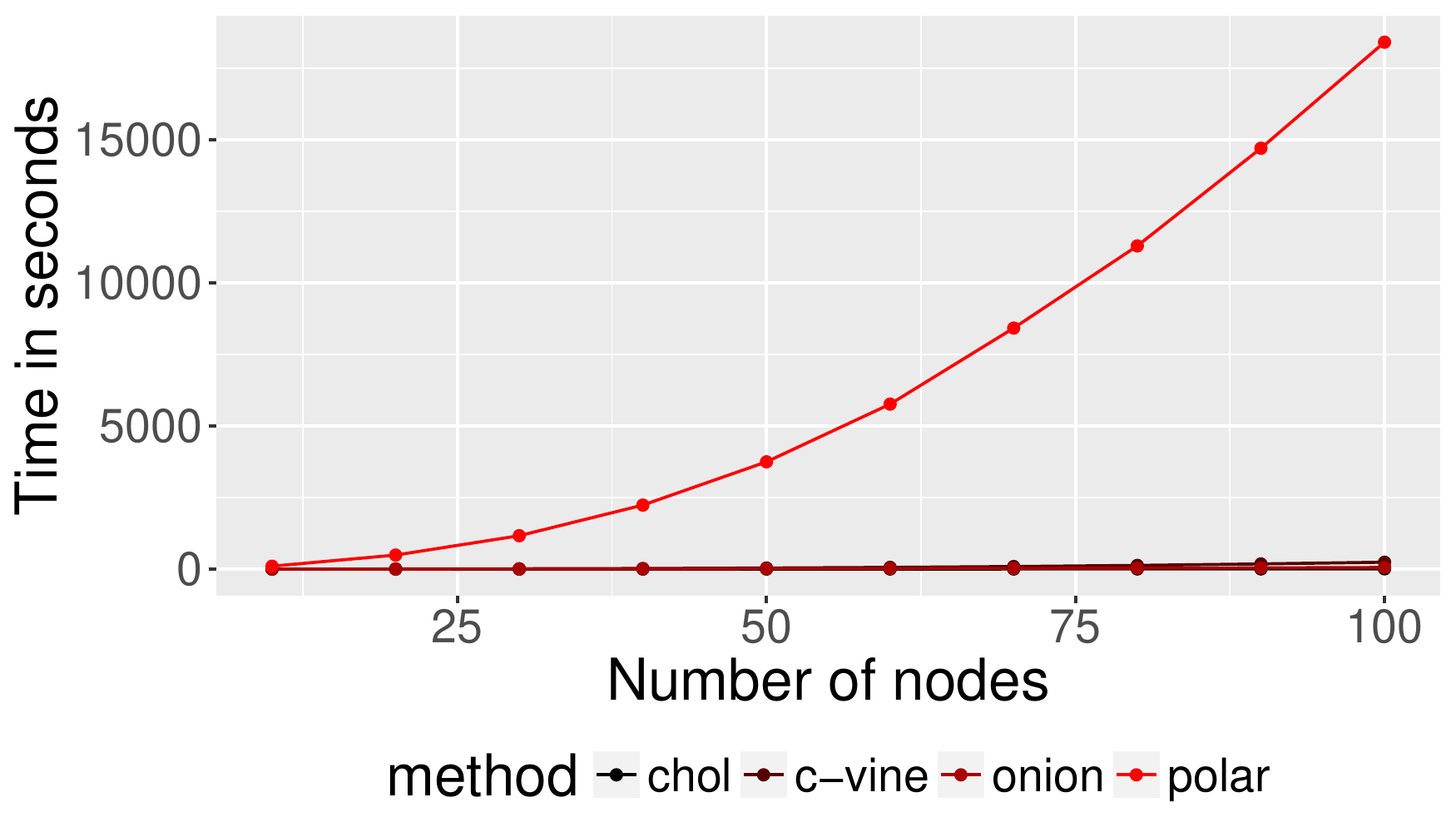}
	\includegraphics[width = 0.45\textwidth]{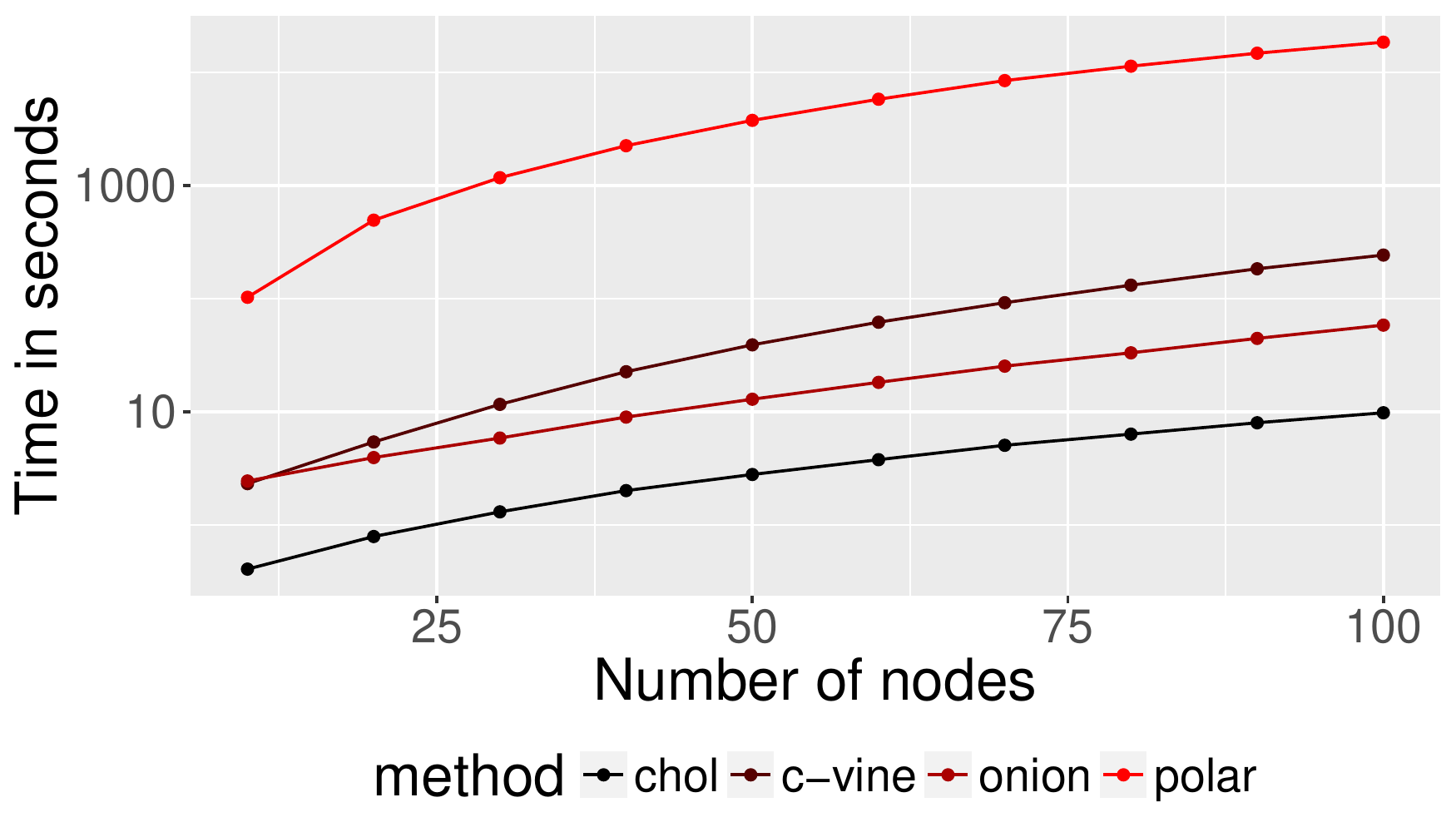}
	\caption{Execution time of available methods for uniform sampling of
		correlation matrices, both in linear (left) and logarithmic (right)
		scale. \texttt{chol}: our proposal; \texttt{c-vine},
		\texttt{onion}: methods by \cite{lewandowski2009}; \texttt{polar}: method by \cite{pourahmadi2015}.} 
	\label{fig:time}
\end{figure}

The results of the experiment are shown in Figure~\ref{fig:time}. We observe
that our method is faster than all of the existing approaches in the literature.
The polar parametrization method has the worst performance, several orders of
magnitude slower than the other algorithms. This can be explained by the use
of inverse transformation sampling for simulating the angles. By contrast, our
method achieves highly competitive results by taking advantage of the direct
representation provided by the Cholesky factorization, as well as the simple
form of the target distribution and the proposed values on each iteration. The
scripts used for generating the data and figures described throughout the paper
are publicly available, as well as the implementation of the algorithms
described\footnote{\url{https://github.com/irenecrsn/rcor}}, so all the above
experiments can be replicated. 

\section{Conclusions}\label{sec:conc}
In this paper we have proposed a Metropolis-Hastings method for uniform sampling
of correlation matrices. We have studied its properties, both
theoretically and empirically, and shown fast convergence to the target uniform
distribution. We have also executed a comparative performance study, where our
approach has yielded faster results than all of the related approaches in the
literature.

In the future, we would like to further explore variants of our Markov chain
algorithm, such as the independent Metropolis or adaptive schemes. We would also
like to expand on the theoretical convergence analysis of such variants, as well
extend the empirical convergence monitoring to other relevant quantities apart
from the acceptance ratio.

{\footnotesize
\textbf{Acknowledgements}. This work has been partially supported by the Spanish Ministry of Economy,
Industry and Competitiveness through the Cajal Blue Brain (C080020-09; the
Spanish partner of the EPFL Blue Brain initiative) and TIN2016-79684-P
projects; by the Regional Government of Madrid through the
S2013/ICE-2845-CASI-CAM-CM project; and by Fundación BBVA grants to Scientific
Research Teams in Big Data 2016.
I. Córdoba {has been} supported by the predoctoral grant FPU15/03797
from the Spanish Ministry of Education, Culture and Sports. G. Varando
{has been partially} supported by research grant 13358 from VILLUM FONDEN.}

\bibliographystyle{splncs04}
\bibliography{paper}

\end{document}